# Towards atom counting from first moment STEM images: methodology and possibilities

Yansong Hao[1,2], Annick De Backer[1,2], Scott David Findlay[3], Sandra Van Aert[1,2]

# Abstract

Through a simulation-based study we develop a statistical model-based quantification method for atomic resolution first moment scanning transmission electron microscopy (STEM) images. This method uses the uniformly weighted least squares estimator to determine the unknown structure parameters of the images and to isolate contributions from individual atomic columns. In this way, a quantification of the projected potential per atomic column is achieved. Since the integrated projected potential of an atomic column scales linearly with the number of atoms it contains, it can serve as a basis for atom counting. The performance of atom counting from first moment STEM imaging is compared to that from traditional HAADF STEM in the presence of noise. Through this comparison, we demonstrate the advantage of first moment STEM images to attain more precise atom counts. Finally, we compare the integrated intensities extracted from first-moment images of a wedge-shaped sample to those values from the bulk crystal. The excellent agreement found between these values proves the robustness of using bulk crystal simulations as a reference library. This enables atom counting for samples with different shapes by comparison with these library values.

# Introduction

Over the past decade, STEM has become one of the most powerful tools for the characterization of complex nanomaterials thanks to its ability to acquire images down to sub-nanometer resolution [1–3]. During image acquisition, a sharply-focused electron probe scans across the sample and electrons scattered at certain angles are collected by detectors at each scanning position. The collected electrons are then used to generate images containing features at the level of individual atoms. By selecting an annular detector with desired collection range, various STEM imaging modes are available [4]. HAADF STEM in particular has become a popular technique because it provides images that are directly interpretable. Previous studies have established that the contrast of HAADF STEM images strongly depends on the atomic number and the number of atoms in the atomic columns [5–10].

Nevertheless, visual interpretation of atomic-resolution HAADF STEM images is often insufficient to fully reveal the structural information of a specimen. This limitation hinders a detailed understanding of the structure-property relationship in nanomaterials since their chemical and physical properties are intrinsically determined by their structural and chemical composition. To address this inadequacy, model-based quantification methods for HAADF STEM images have been developed. These methods enable high precision measurements of material structure parameters using statistical parameter estimation theory [11–14]. This theory makes use of a quantitative model which is parametric in the position, intensity and width of the image intensity peaks. By fitting this model to the observed STEM images, the total scattering intensity per atomic column can be calculated. This approach has been used to boost the quantitative analysis of HAADF STEM data and to solve diverse problems in material science.

Successful quantification of image intensities has enabled compositional mapping of crystals at atomic resolution [15,16]. More importantly, it also provides a roadmap to perform atom counting from HAADF STEM images viewed along a zone axis [17,18]. By combining counting results from more than one viewing direction, the 3D atomic arrangement of crystalline nanoparticles was successfully reconstructed with single atom sensitivity [19]. Efforts have been made to achieve 3D reconstructions from a single projection [20–23] since acquiring HAADF STEM images from different viewing directions is not always feasible for nanostructures that are very beam sensitive or are observed in-situ, e.g. under the flow of a gas. Recent examples include the successful reconstruction of a Pt nanoparticle in different gaseous environments [24] and supported Au nanoparticles at high temperature [25]. However, the nature of HAADF STEM, which relies on high-angle scattered electrons for imaging, imposes certain limitations. Firstly, it is not suitable for weakly-scattering samples [26]. Secondly, because the atomic column intensities in HAADF STEM scale with the square of the atomic number [27], light columns are easily obscured by the stronger intensity of heavier columns, making it challenging to simultaneously estimate the structural parameters for both light and heavy atomic columns.

Recent advances in the design of pixelated electron detectors have introduced new possibilities to overcome these limitations. A pixelated detector contains a large number of pixels, where each pixel works as an individual detector. Instead of collecting electrons only within a certain range of scattering angles, a pixelated detector can record the full convergent beam electron diffraction (CBED) pattern at each scanning position, resulting in a 4D dataset (2D CBED patterns on a 2D scanning grid). Such datasets offer significant flexibility in generating diverse STEM imaging modes, either by feeding the datasets into computational reconstruction algorithms or applying virtual detectors to each CBED pattern [28,29]. A representative for the former type of method is ptychography, recently demonstrated to be able to retrieve information about the sample along its thickness direction [30,31], but its computational complexity is considerable and its robustness for atom counting, particularly at lower doses, remains unclear. A notable example of the latter type is first-moment STEM imaging which describes the center-of-mass (COM) shift within the CBED patterns at each scanning position. First moment imaging has been used to solve versatile problems in materials science, for instance, local mapping of electric fields [32,33], visualization of magnetic structures [34], measurement of magnetic domains [35], imaging of 2D materials [36] and determination of charge densities at the atomic scale [37]. Given model-based parameterization has proven fruitful for quantitative HAADF STEM imaging, as discussed earlier, it is worth exploring whether a similar model-based parameterization is possible for first moment STEM images. Within the phase object approximation, the first moment STEM images are proportional to the gradient of the atomic column projected potential convolved with the probe intensity, which scales linearly with both the number of atoms in the atomic columns and the atomic number [33,36,38,39]. As compared to HAADF STEM imaging, this reduces the signal difference between light and heavy atomic columns, potentially enabling parameter-estimation-based atom counting to be extended to samples with both heavy element columns and light element columns present.

In this work, we develop a model-based quantification method for first moment STEM images, highlighting the advantages of using this imaging mode for atom counting. Our method starts with fitting a parametric model to simulated COM images. This enables the determination of structure parameters of the images, including the width, height and position of an atomic column. The integrated intensity per atomic column is then calculated from these parameters, and its response is studied as a function of thickness. In this manner, the possibility of using COM images for atom counting is thoroughly examined. Its precision is further compared to HAADF STEM imaging in the presence of noise. Finally, the integrated intensity from a wedge-shaped specimen is compared to values obtained from bulk crystal in simulation, demonstrating the robustness of our method for atom counting in specimens of varying shapes.

# Methods

## *4D-STEM simulation and synthesis of COMx(y) images*

Using aluminum as an example material, 4D STEM datasets of a crystal viewed along the [001] zone axis were simulated up to a thickness of 40 atoms, i.e. about 16 nm. All simulations were performed with the MULTEM software which uses the multislice algorithm within the frozen-phonon framework [40,41]. For each simulation, the defocus of the incoming electron probe is set to half the overall thickness, providing optimal contrast for the COM images [29,39,42]. Additional simulation details are provided in **Table 1**. An aberration-free probe is assumed, being an appropriate approximation for an aberration-corrected probe whose aberrations are well-balanced in the probe-forming aperture.

Table 1 Settings used for the multislice simulations in the MULTEM software

| Parameter | Value |
|---|---|
| Acceleration voltage | 300 kV |
| Defocus | Half of overall thickness |
| Spherical aberration | 0 mm |
| Semi-convergence angle | 20 mrad |
| FWHM of the source spatial incoherence | 0.5 Å |
| Number of unit cells per supercell | 7×7 |
| Pixel size in real space | 0.2025 Å |
| Number of pixels in real space | 20×20 |
| Pixel size in reciprocal space | 0.0353 Å$^{-1}$ |
| Number of pixels in reciprocal space | 180×180 |
| Maximum sampling angle in reciprocal space | 62.5 mrad |
| Root mean square displacement Al atom | 0.085 Å |

The 4D dataset for a wedge-shaped aluminum structure in the [001] zone axis was simulated under the same set of microscope parameters (acceleration voltage, aberrations and semi-convergence angle). More detailed simulation parameters are provided in **Table S1**. **Figure 1** shows the input structure of the aluminum wedge (left) and a map of the number of atoms in each atomic column (right).

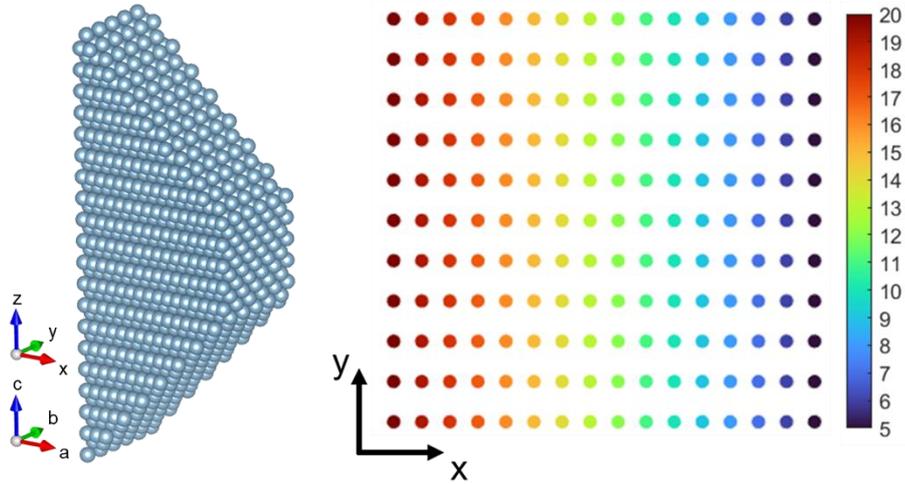

**Figure 1. Input structure of the aluminum wedge as a 3D model (left) and map of the number of atoms in each column (right).**

The simulated 4D dataset for the crystal and the wedge were then convolved with a 2D Gaussian function to account for the spatial incoherence of the probe. In practice, the COM shifting is typically determined along two perpendicular directions, generating COMx and COMy images. To synthesize these images, each CBED in the datasets was multiplied by the virtual first moment STEM detectors as shown in **Figure 2,** corresponding to the $[k_x; k_y]$ axes in reciprocal space, and followed by a summation over all entries.

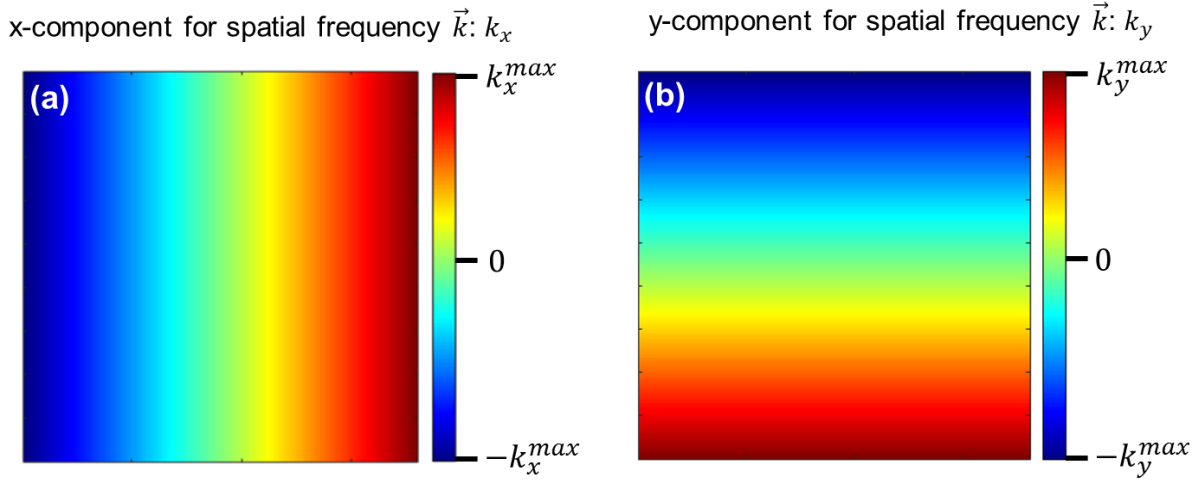

**Figure 2. Virtual detectors to synthesize (a) COMx and (b) COMy image from a 4D dataset.**

Mathematically, the construction is described by the following equation:

$$I^{kl}_{COMx} = \sum_{p=1}^{P}\sum_{q=1}^{Q} k_x^{pq} I_{pqkl} \;;\; I^{kl}_{COMy} = \sum_{p=1}^{P}\sum_{q=1}^{Q} k_y^{pq} I_{pqkl} \qquad Eq(1)$$

where $I_{pqkl}$ refers to the intensity of the CBED patterns at reciprocal pixel $(k_p, k_q)$ at probe position $(x_k, y_l)$. $k_x^{pq}$ and $k_y^{pq}$ are the x- and y- components of the spatial frequency at reciprocal pixel $(k_p, k_q)$, respectively. $I^{kl}_{COMx}$ and $I^{kl}_{COMy}$ are the image intensity of COMx and COMy image at pixel (k, l) corresponding to probe position $(x_k, y_l)$.

## Model-based parameter estimation

To precisely determine quantities such as atomic column positions, peak intensities and integrated intensities from atomic-resolution STEM images, it can be fruitful to consider images as data planes from which those parameters are measured using statistical parameter estimation theory [13]. This procedure begins with constructing a parametric model, which depends on all unknown structure parameters. This model describes the expectations of the pixel values in STEM images. By fitting this model to the images using a criterion of goodness of fit, the unknown structure parameters can be determined. For first moment STEM imaging, the COM shift vector of each CBED is calculated, which in the phase object approximation is proportional to the gradient of the projected potential of atomic columns convolved with the probe intensity. Since the projected potential $P(r,\theta)$ of a crystal is peaked at the atomic column positions [43,44] and the probe intensity should be rotationally symmetric, we will describe the projected potential convolved with the probe intensity as a superposition of 2D Lorentzian peaks:

$$P(r,\theta) = \sum_{m=1}^{M} \eta_m \left(\rho_m^2 + (x - \beta_{x_m})^2 + (y - \beta_{y_m})^2\right)^{-\frac{3}{2}} \qquad Eq(2)$$

with $M$ the total number of atomic columns, $r = (x, y)^T$ corresponding to the STEM probe position and $\theta = (\beta_{x_1}, \ldots, \beta_{x_M}, \beta_{y_1}, \ldots, \beta_{y_M}, \rho_1, \ldots, \rho_M, \eta_1, \ldots, \eta_M)^T$ representing the vector of unknow parameters. COMx(y) images are accordingly modelled as a superposition of the X(Y) derivatives of this 2D Lorentzian function. The expectation model $f_{kl}(\theta)$ for the COMx image, giving the expectation value of pixel $(k, l)$ at position $(x_k, y_l)$, is then written as:

$$f_{kl}^{COMx}(\theta) = \sum_{m=1}^{M} -3\,\eta_m (x_k - \beta_{x_m})(\rho_m^2 + (x_k - \beta_{x_m})^2 + (y_l - \beta_{y_m})^2)^{-\frac{5}{2}} \qquad Eq(3)$$

where $\frac{\eta_m}{\rho_m^3}$, $\rho_m$, $\beta_{x_m}$ and $\beta_{y_m}$ are the height, width, x- and y- coordinates of the 2D Lorentzian associated with the $m$th atomic column (the model for COMy follows trivially). The unknown parameters of the original 2D Lorentzian peaks given by Eq(2) are found by fitting its X(Y) derivative given in Eq(3) to the COMx(y) image. The integrated scattering intensity from the $m$th atomic column is then calculated from the estimated width and height:

$$V_m = 2\pi \frac{\eta_m}{\rho_m^3} \rho_m^2 = 2\pi \frac{\eta_m}{\rho_m} \qquad Eq(4)$$

The advantage of using a 2D Lorentzian-based model to describe the projected potential convolved with the probe intensity, as compared to the more commonly used 2D Gaussian-based model, is discussed later in this paper.

Since the expectation models describing the COMx and COMy images share the same set of unknown parameters, these models are fitted simultaneously to the COMx and COMy images, hence resulting in a higher precision for the estimated parameters as compared to estimating the parameters from the COMx or COMy images independently. Use is made of the uniformly weighted least squares criterion, which evaluates the correspondence between the image and the model for both COMx and COMy. The estimates $\hat{\theta}$ are given by the values of $t$ that minimize the uniformly weighted least square criterion:

$$\hat{\theta} = \arg\min_{t} \sum_{k=1}^{K}\sum_{l=1}^{L}\left[\left(w_{kl}^{COMx} - f_{kl}^{COMx}(t)\right)^2 + \left(w_{kl}^{COMy} - f_{kl}^{COMy}(t)\right)^2\right] \qquad Eq(5)$$

where $w_{kl}^{COMx}$ and $w_{kl}^{COMy}$ represent the observed values in the COMx and COMy images at pixel $(k, l)$. Direct implementation of this criterion for images containing a large number of atomic columns is computationally expensive. Therefore, an efficient model estimation algorithm introduced in [45] has been used.

# Results

## *Atom counting for bulk crystals*

As described in the previous section, 4D STEM datasets are simulated for aluminum in the [001] zone axis up to a thickness of 40 atoms. Examples of simulated COMx and COMy images, and the corresponding models evaluated at the estimated parameters, are shown in **Figure 3(a-d)**. The appearance of COM images can be described as follows [33,38,39]. When the incoming electrons pass through the material, they encounter a net force of attraction towards the partially-screened nuclei. Hence, the electron probe is deflected to the right as it passes to the left of an atomic column resulting in more electrons being scattered to the positive region of the virtual detector when considering COMx imaging (see **Figure 2**). The opposite occurs when the probe passes to the right of a column. As a result, a contrast reversal is observed as the probe scans across an atomic column, as seen in **Figure 3(a, c)**.

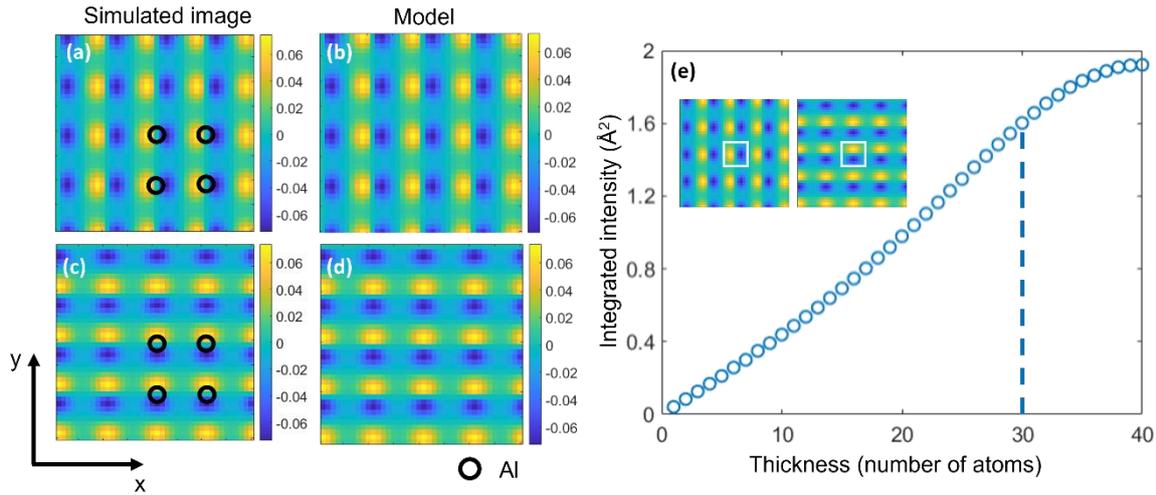

**Figure 3.** Central part of simulated (a) COMx and (c) COMy image for an aluminum crystal of 5 atoms thick with the corresponding model shown in (b) and (d), respectively. (e) Integrated intensity of the central atomic column as a function of thickness.

For both COMx and COMy, the model (**Figure 3(b, d)**) agrees well with the simulated image (**Figure 3(a, c)**) in terms of contrast features and the absolute signal range, demonstrating the feasibility of fitting a superposition of x(y)- derivative of 2D Lorentzians to COMx(y) images. **Figure 3(e)** shows the dependence of the integrated intensity (Eq(4)) as a function of thickness for the central atomic column. A monotonic increase is observed up to approximately 35 atoms thickness (~14nm) before the value levels off. This range is narrower compared to HAADF STEM, where the scattering cross section increases (almost) linearly up to much larger thicknesses [46,47]. The phase-object approximation predicts the COMx(y) image to be related to the gradient of the projected potential of atomic columns, which would vary linearly with the number of atoms in the column, but the domain of validity of this approximation is limited. Previous work suggests that the breakdown of this approximation first manifests as a quantitative underestimate for the projected potential [39,48]. In our parametric model, the projected potential of an atomic column is described by the integrated intensity of its Lorentzian peak, and so the underestimation manifests as a saturation of the integrated intensities at large thickness, instead of a continuous increase, as seen in **Figure3(e)**. (Aluminum is a relatively light scatterer; the phase object approximation may break down sooner in more strongly scattering samples.) Despite such saturation, the monotonic dependence of the integrated intensity on thickness demonstrates its utility as a quantitative measure to count the number of atoms from COMx and COMy images, especially for thin samples.

To explore the potential advantages of using COMx and COMy images for atom counting, the precision with which the integrated intensity can be measured is compared between COM and traditional HAADF STEM. For this purpose, repetitive noise realizations are performed for both imaging modes. The procedure to simulate noise realizations is summarized briefly below. In practice, the reciprocal pixels in the 4D datasets represent statistically independent electron counting results [49]. For a finite incident electron dose, the expectation model including the electron dose at reciprocal pixel $(k_p, k_q)$ at probe position $(x_k, y_l)$ equals:

$$\lambda_{pqkl} = N_d \Delta x \Delta y I_{pqkl} \qquad Eq(6)$$

where $N_d$ is the incident electron dose ($e^-/\text{Å}^2$), $\Delta x = \Delta y$ is the pixel size, and $I_{pqkl}$ is the intensity of the wavefield at $(k_p, k_q)$ (normalized to sum to unity across the CBED pattern such that, as per the Born interpretation, it gives the probability for the electron to be recorded in that detector pixel). Due to the inevitable presence of electron counting noise in real experiments, the actual pixel values fluctuate randomly and are considered to be Poisson distributed with expectation value $\lambda_{pqkl}$. For each thickness in the 4D STEM dataset, 100 noise realizations are simulated at two dose levels ($N_d = 5 \times 10^5\ e^-/\text{Å}^2$, and $N_d = 1 \times 10^5\ e^-/\text{Å}^2$), followed by a normalization with respect to the incident electron dose (i.e. dividing each noise realization by the incident electron dose per pixel). This normalization places the noise realizations on the same intensity scale as the raw 4D STEM simulations ($I_{pqkl}$) hence allowing a direct comparison between them. COMx and COMy images are then synthesized from the simulated 4D dataset with Poisson distributed noise. Examples of noise realizations for the COMx image are shown in **Figure 4** together with the corresponding image synthesized directly from the simulated 4D dataset (in the absence of Poisson distributed noise).

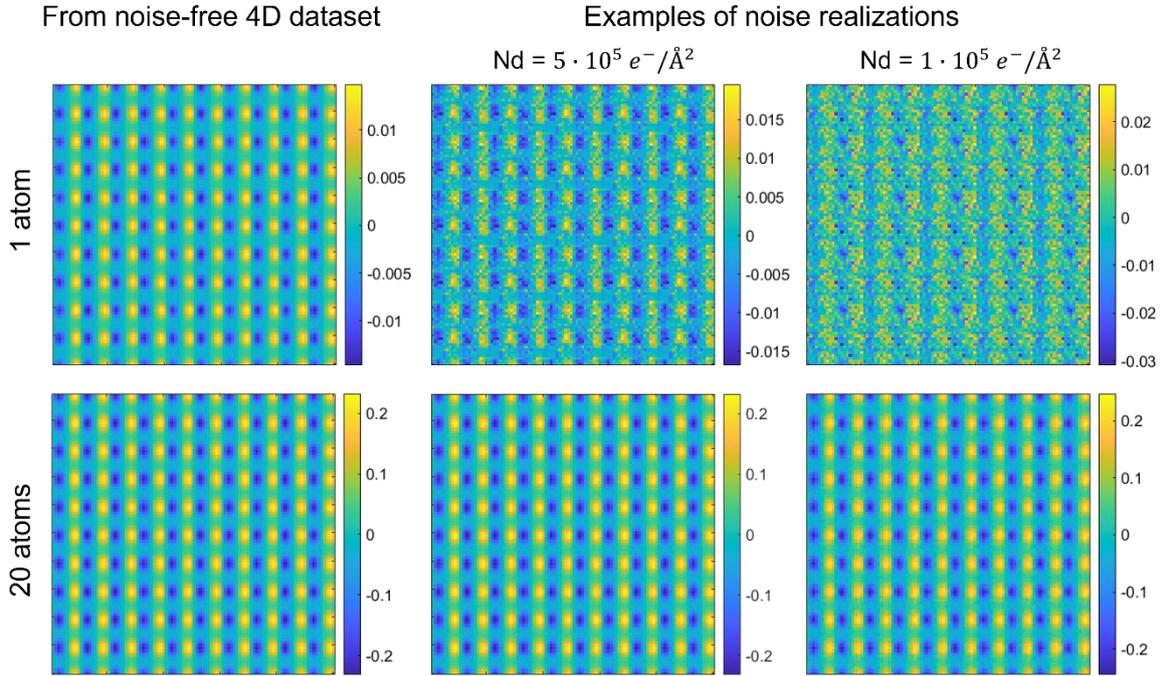

**Figure 4.** COMx images computed from the noise-free simulated 4D dataset for a thickness of 1 and 20 atoms, together with examples noise realizations at $N_d = 5 \times 10^5$ e$^-$/Å$^2$ and $N_d = 1 \times 10^5$ e$^-$/Å$^2$.

For each noise realization, the unknown parameters of Eq(3) have been estimated, which enables us to evaluate the precision of the estimation. **Figure 5(a, b)** show the distribution of the integrated intensities of the central atomic column for thicknesses between 5 and 20 atoms. For comparison, HAADF STEM images of the same aluminum crystal thickness series have been simulated under the same set of microscope parameters (see **Table S2** for simulation details), with noise realizations generated using a similar strategy and integrated intensities estimated by the StatSTEM software [45]. The results are shown in **Figure 5(c, d)**. In the absence of noise, these histograms would contain isolated components, with each component corresponding to atomic columns having the same number of atoms. However, electron counting noise as accounted for here, and other factors that exist in experiments such as instability of the material under the electron beam and intensity transfer between neighboring columns, can lead to a smearing out of the components [17]. Overlap between neighboring components would negatively affect the accurate correlation between the integrated intensity and the number of atoms in a column, thus increasing the probability of miscounting the number of atoms.

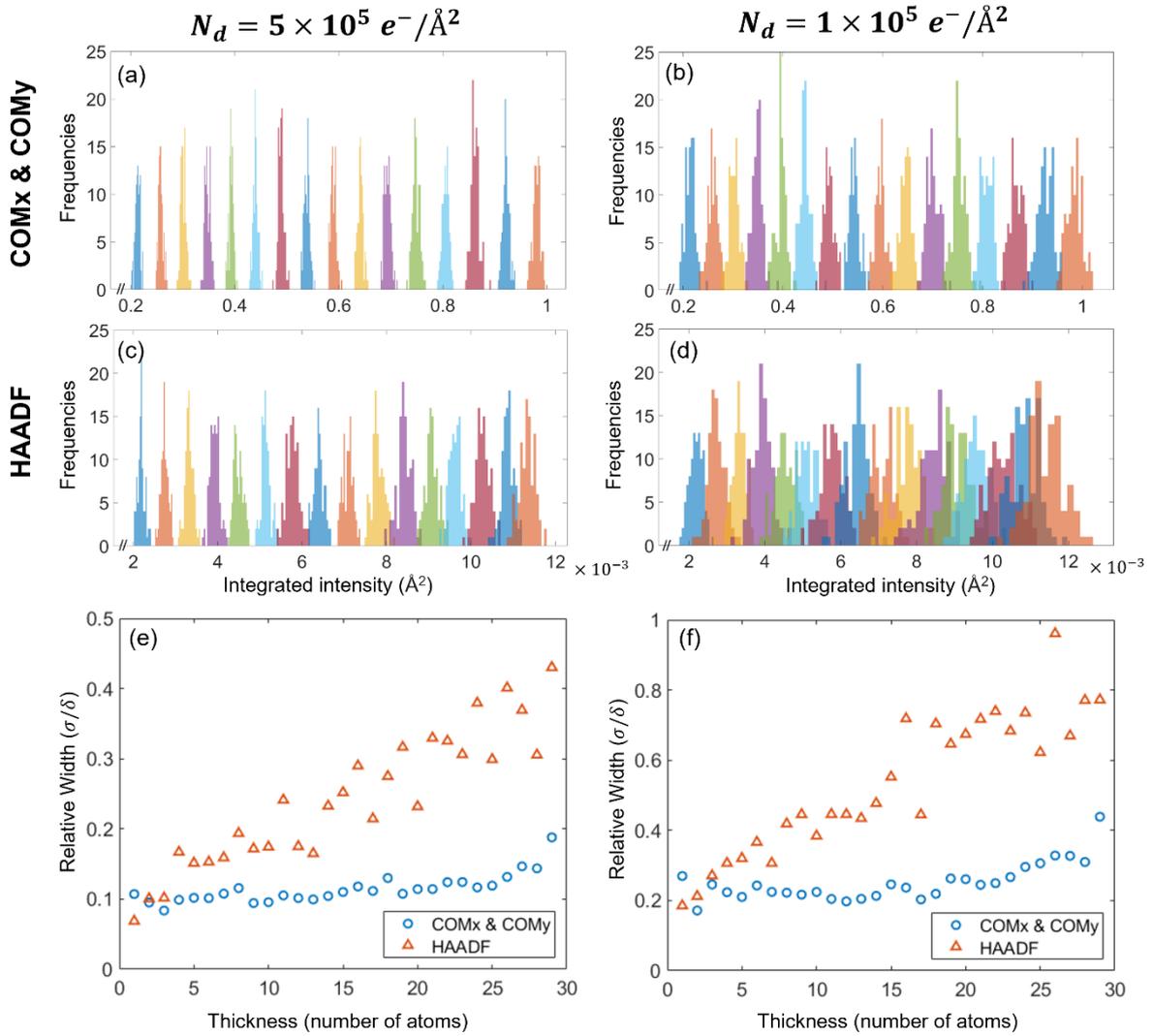

**Figure 5.** Analysis of the estimated integrated intensity from repetitive noise realizations at the two dose levels $N_d = 5 \times 10^5 \ e^-/\text{Å}^2$ (left column) and $N_d = 1 \times 10^5 \ e^-/\text{Å}^2$ (right column). Histograms of the estimated integrated intensities from (a, b) COMx & COMy and (c, d) HAADF STEM. (e, f) Relative width as a function of thickness.

Although smeared out, well-separated components are observed for COMx and COMy imaging if the incident electron dose is sufficiently large, with each component generated by atomic columns of the same thickness (i.e. containing the same number of atoms). When the dose is reduced from $5 \times 10^5$ to $1 \times 10^5$ $e^-/\text{Å}^2$, slight overlap between neighboring components appears, but individual components can still be distinguished. This differs from HAADF-STEM imaging, where overlap between neighboring components cannot be neglected even at higher incident electron dose (**Figure 5(c)**) and becomes more pronounced with increasing integrated intensity, i.e. with increasing number of atoms in the column. Reducing the incident electron dose worsens the situation, as illustrated by the significant overlap in **Figure 5(d).** In this case, it is no longer possible to visually distinguish individual components in the histogram.

The qualitative observation of less overlap between neighboring components implies atom counting will be more precise when using COMx and COMy images as compared to HAADF STEM. This reduced overlap can be quantified by the relative width, which is the ratio between the standard deviation of each component ($\sigma$) and the increment between consecutive components ($\delta$). The relative width as a function of thickness for both imaging modes is shown in **Figure 5(e, f)**. At both dose levels, the relative width of COMx and COMy imaging is on average almost three times lower than that of HAADF STEM images, highlighting the benefit of using COMx and COMy images to attain more precise atom counting. Moreover, the relative width for HAADF STEM imaging scales significantly with thickness, reflecting the more significant overlap between components with increasing integrated intensity as observed in **Figure 5(c, d)**. This will result in less precise atom counts for higher thicknesses in HAADF STEM. In contrast, the relative width for COMx and COMy images is less sensitive to thickness since the overlap between neighboring components is more constant. This ensures constant precision in atom counting regardless of increasing thickness. This can be understood as follows. In HAADF STEM, the integrated intensity of an atomic column is proportional to the total number of electrons scatted to the high-angle detector by the column. With increasing thickness, more electrons are scattered to the detector and so the standard deviation of the integrated intensity increases since the electron counting process is governed by Poisson statistics (see Eq(6)). Meanwhile, the increment of the integrated intensity between columns having $n$ and $n + 1$ atoms remains constant since each atom contributes in an approximately equal and independent way. Consequently, the relative width scales as a function of thickness. In COMx and COMy imaging, the increment of the integrated intensity between columns is again approximately constant since the integrated intensity is proportional to the projected potential of the atomic column which scales linearly with the number of atoms in the column. However, in contrast to HAADF STEM, the standard deviation of the integrated intensity is now also approximately constant since the total number of electrons used to form the COMx and COMy images is essentially the same at all thicknesses. Thus, the relative width is approximately constant.

## *Atom counting for a wedge-shaped sample*

By using bulk crystals, our model-based approach for COMx(y) images has been shown to successfully isolate contribution from individual atomic columns, forming a basis for column-by-column atom counting. However, this is insufficient to demonstrate the applicability of this approach for samples where adjacent columns have different number of atoms. In this section, we show the capability of counting atoms from COMx and COMy images for a more realistic wedge-shaped sample, which is considered as a prototype for catalyst nanoparticles. The MULTEM software has been used to simulate the 4D dataset for an aluminum wedge with a thickness ranging from 5 to 20 atoms. The microscope parameters for this simulation are the same as those listed in **Table 1** (see **Table S1** for more simulation details).

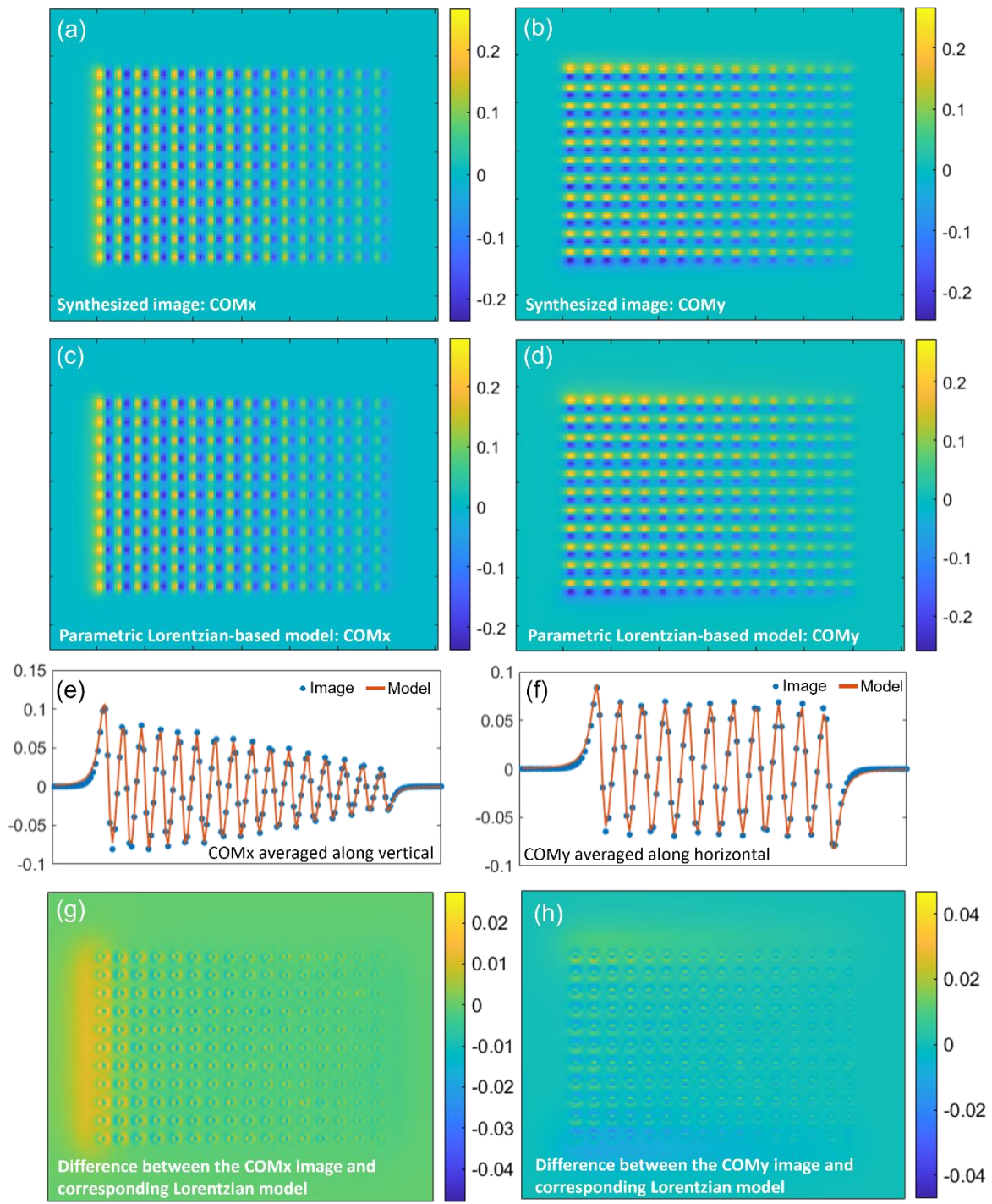

**Figure 6. (a, b) Synthesized COMx and COMy images of the aluminum wedge, and (c, d) corresponding parametric Lorentzian-based model. (e) Line profile for the synthesized COMx image and the Lorentzian model averaged along the vertical direction. (f) Line profile for synthesized COMy image and the Lorentzian model averaged along the horizontal direction. (g, h) Difference between COMx(y) image and corresponding Lorentzian model.**

**Figure 6(a, b)** shows the COMx and COMy images simulated for this structure and **Figure 6(c, d)** shows the parametric models, given by Eq(3) evaluated at the parameters estimated following Eq(5). Visual inspection reveals no significant difference between the models and the synthesized images. Contrast features are successfully reproduced by the model, with stronger signals for atomic columns containing more atoms. The excellent fit is demonstrated further in **Figure 6(e, f)**, where the pixel values of the images and models are compared after averaging along the vertical direction for COMx and along the horizontal direction for COMy. A full difference map between the synthesized image and the corresponding model is also shown in **Figure 6(g, h),** evidencing an accurate quantification of the COMx and COMy image. The estimated integrated intensities of the atomic columns are then grouped according to the number of atoms in each column. **Figure 7** compares the average intensity of each group to the values estimated from the aluminum bulk crystal at the corresponding thickness, showing good agreement**.** This consistency demonstrates the successful isolation of individual atomic column contributions in the wedge-shaped sample and highlights the robustness of using COMx and COMy images for atom counting, where bulk crystal simulations can serve as a reference library. Consequently, atom counting for samples with different shapes can be achieved by comparing to these library values.

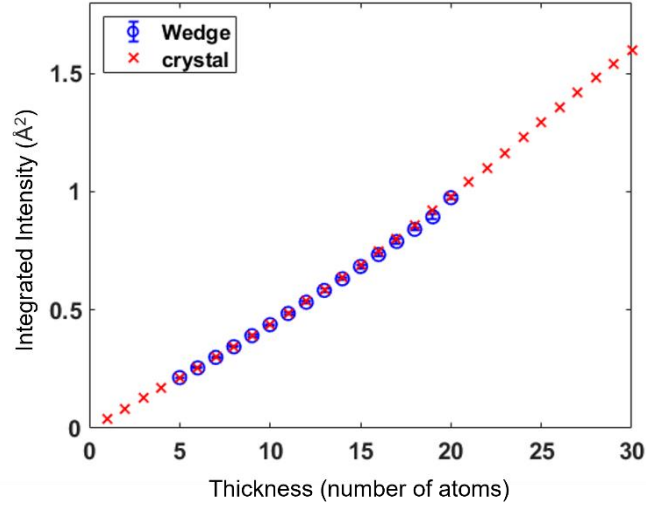

Figure 7. Integrated intensity comparison between the aluminum wedge and the aluminum bulk crystal.

# Discussion

So far, we have successfully developed a model-based approach to quantify COMx and COMy images synthesized from simulated 4D STEM datasets. This allows for a more precise atom counting as compared to the traditional HAADF STEM imaging. For the suggested parametric model, the projected potential convolved with the probe intensity is approximated by a superposition of 2D Lorentzians (Eq2), with localized peaks at each atomic column. Unknown quantities, such as the height, width and volumes of these peaks, are then determined by simultaneously fitting the COMx(y) images as a superposition of x(y)- derivatives of 2D Lorentzians. This differs from the previous use of a 2D Gaussian to approximate the projected potential [43] and HAADF STEM images [45,47,50–52]. In this section we therefore compare the use of a superposition of x(y)- derivatives of 2D Lorentzians with that of a superposition of x(y)- derivatives of 2D Gaussians to show that the former is better for parameterizing the COMx(y) images.

Taking COMx as an example, the parametric model describing the expectations of the image intensities at pixel (k, l) using x(y)- derivatives of 2D Gaussians is given by:

$$f_{kl}(\theta) = \sum_{i=m}^{M} -2\,\eta_m \frac{(x_k - \beta_{x_m})}{2\rho_m^2} \exp\left[\frac{-(x_k - \beta_{x_m})^2 - (y_l - \beta_{y_m})^2}{2\rho_m^2}\right] \quad Eq(7)$$

where $\eta_m$, $\rho_m$, $\beta_{x_m}$ and $\beta_{y_m}$ are the height, width, x- and y-coordinates of the 2D Gaussian associated with the $m$th atomic column. To compare the quality of Lorentzian- and Gaussian-based models, the synthesized COMx and COMy images of the aluminum wedge were parameterized following Eq(7) and the results are shown in **Figure 8.**

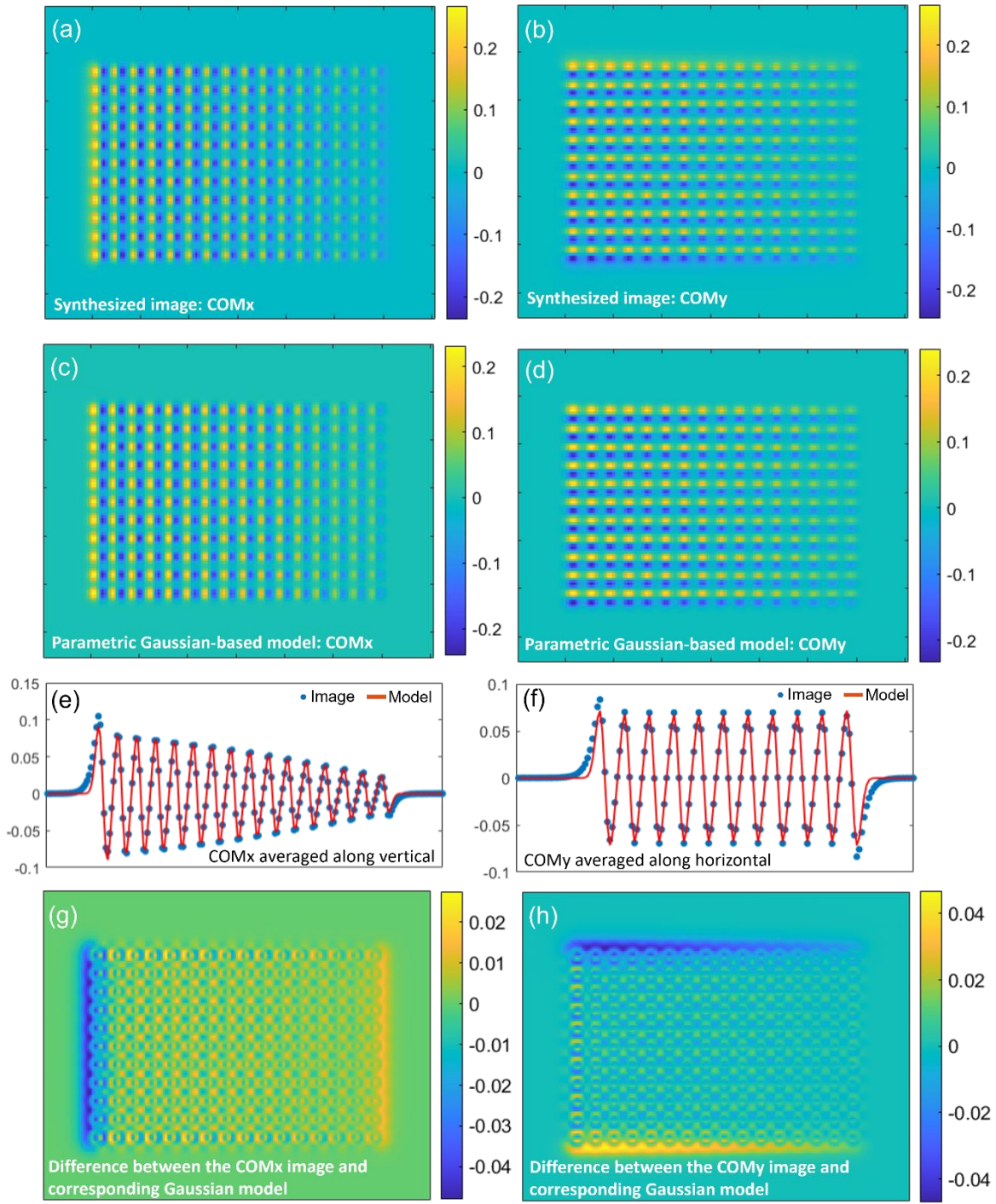

**Figure 8.** (a, b) Synthesized COMx and COMy images of the aluminum wedge, and (c, d) corresponding parametric Gaussian-based model. (e) Line profile for the synthesized COMx image and the Gaussian model averaged along the vertical direction. (f) Line profile for synthesized COMy image and the Gaussian model averaged along the horizontal direction. (g, h) Difference between COMx(y) image and corresponding Gaussian model.

Qualitatively, a Gaussian-based model can still capture the contrast features of the synthesized images. However, the level of quantification is not as accurate as when a Lorentzian-based model is used, which is evidenced by the larger deviations between the synthesized image and the model, as illustrated in **Figure 8(e-h)** (see **Figure 6(e-h)** for a comparison). To further compare the quality of the fit in a quantitative manner, the sum of squared residuals (SSR) is calculated for both models:

$$SSR = \sum (w_{kl}^{COMx(y)} - f_{kl}^{COMx(y)}(\hat{\theta}))^2 \qquad Eq(8)$$

where $w_{kl}^{COMx(y)}$ is the value of pixel (k, l) in COMx(y) image and $f_{kl}^{COMx(y)}(\hat{\theta})$ is the corresponding pixel value given by the Lorentzian- or the Gaussian- based parametric model. The results are listed in **Table 2**. Consistent with the qualitative observation, a Lorentzian-based model provides a significantly more accurate quantification of the COMx and COMy images, as indicated by its nearly five times lower SSR value.

Table 2 SSR of Lorentzian- and Gaussian- based parametric model for COMx(y) image

| Parametric model | COMx | COMy |
|---|---|---|
| Lorentzian-based | 0.40 | 0.41 |
| Gaussian-based | 1.89 | 2.30 |

The superior quality of the Lorentzian-based model is likely related to its more accurate description of the projected potential of atomic columns convolved with the probe intensity. To validate this claim, the projected potential of a 10-atom-thick aluminum column was simulated and convolved with the probe intensity:

$$O(r) = P(r) * psf(r) \qquad Eq(9)$$

where $P(r)$ is the projected potential and $psf(r)$ is the probe intensity. **Figure 9** shows the results of fitting both Lorentzian (Eq(2)) and Gaussian models (see section 2 in supporting info for more details) to $O(r)$. Clearly, the Lorentzian model leads to a more accurate parameterization, which is consistent with the better performance of Lorentzian-based model for COMx and COMy images of the aluminum wedge. That this parameterization works well for the potential convolved with the probe intensity eliminates the additional complexity of explicitly including the probe intensity in the model, reducing the computational cost of the fitting process.

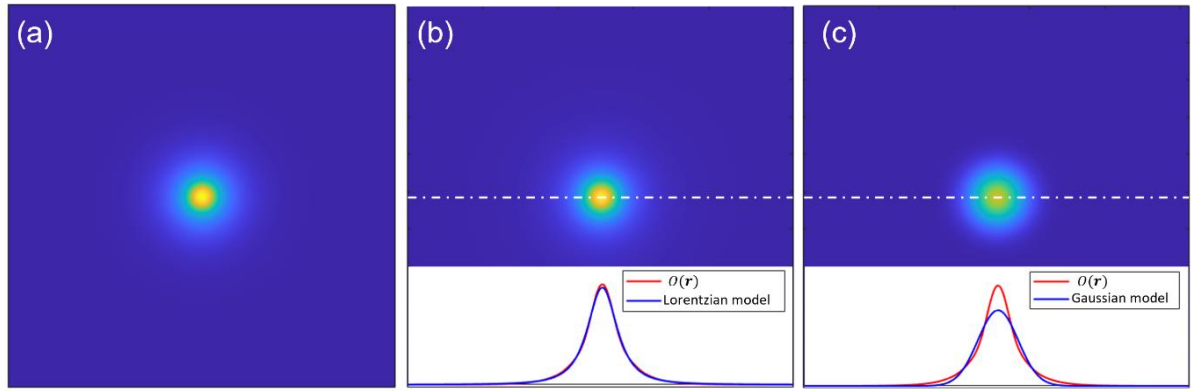

**Figure 9.** (a) Projected potential of 10-atoms-thick aluminum column convolved with the probe intensity. (b) Best fit Lorentzian and (c) Gaussian models illustrated in the same color scale, together with line profiles across the middle of the image.

# Conclusions

In this study, a model-based quantification approach has been developed for first-moment STEM images. By simultaneously fitting COMx(y) images as a superposition of x(y)-derivatives of a 2D Lorentzian peak, contributions from individual atomic columns are successfully isolated, which forms a basis for column-by-column atom counting. Unknown structural parameters, particularly the integrated intensities of atomic columns, are estimated accurately and precisely. The monotonic increase of the integrated intensity as a function of thickness enables the use of this quantitative measure for atom counting. Using HAADF STEM as a reference, the advantage of attaining more precise atom counting from COMx and COMy images is demonstrated by evaluating the relative width of the integrated intensities for both imaging modes. Finally, the robustness of this quantification approach to perform atom counting for samples with different shapes is illustrated. This is achieved by matching integrated intensities from the target sample to library values from bulk crystals.

# Acknowledgements

The authors acknowledge financial support from the Research Foundation Flanders (FWO, Belgium) and Fonds de la Recherche Scientifique (FNRS, Belgium) within the CHISUB network of the EOS (Excellence of Science) program (grant number 40007495). This work was supported by the European Research Council (Grant 770887 PICOMETRICS to SVA). The authors acknowledge financial support from the Research Foundation Flanders (FWO Belgium) through project fundings (G034621N and G0A7723N). The authors acknowledge Dr. Z. Zhang for useful discussions. This research was supported under the Discovery Projects funding scheme of the Australian Research Council (Project No. FT190100619).

# Supporting information

# Towards atom counting from first moment STEM images: methodologies and possibilities

Yansong Hao[1,2], Annick De Backer[1,2], Scott David Findlay[3], Sandra Van Aert[1,2]

*1 Electron Microscopy for Materials Science (EMAT), University of Antwerp, Groenenborgerlaan 171, 2020 Antwerp, Belgium*

*2 NANOlab Center of Excellence, University of Antwerp, Groenenborgerlaan 171, 2020 Antwerp, Belgium*

*3 School of Physics and Astronomy, Monash University, Clayton, Victoria 3800, Australia*

## Section 1: Additional simulation details

**Table S1 Parameters for simulating 4D datasets of the aluminum wedge**

| Parameter | Value |
| --- | --- |
| Acceleration voltage | 300 kV |
| Defocus | Half of overall thickness |
| Spherical aberration | 0 mm |
| Semi-convergence angle | 20 mrad |
| Spatial incoherence of source | 0.5 Å |
| Pixel size in real space | 0.2025 Å |
| Number of pixels in real space | 170×211 |
| Pixel size in reciprocal space | 0.0235 Å$^{-1}$ |
| Number of pixels in reciprocal space | 800×800 |
| Maximum sampling angle in reciprocal space | 185.2 mrad |
| Root mean square displacement Al atom | 0.085 Å |

Table S2 Parameters for simulating HAADF images of the aluminum crystal thickness series

| Parameter | Value |
|---|---|
| Acceleration voltage | 300 kV |
| Defocus | Half of overall thickness |
| Spherical aberration | 0 mm |
| Semi-convergence angle | 20 mrad |
| Spatial incoherence of source | 0.5 Å |
| Number of unit cells per supercell | 7×7 |
| Pixel size in real space | 0.2025 Å |
| Number of pixels in real space | 20×20 |
| Pixel size in reciprocal space | 0.0353 Å$^{-1}$ |
| Number of pixels in reciprocal space | 600×600 |
| Maximum sampling angle in reciprocal space | 208.3 mrad |
| HAADF inner collection angle | 56 mrad |
| HAADF outer collection angle | 190 mrad |
| Root mean square displacement Al atom | 0.085 Å |

# Section 2: Details for simulating and quantifying projected potential of a single aluminum column

Projected potential $P(\boldsymbol{r})$ of a 10-atoms-thick aluminum column was simulated using MULTEM software. The probe intensity $psf(\boldsymbol{r})$ is generated as follows:

$$psf(\boldsymbol{r}) = |p(\boldsymbol{r})|^2$$

where the function $p(\boldsymbol{r})$ is given by the inverse Fourier transform of the objective lens transfer function $T(\boldsymbol{g})$:

$$p(\boldsymbol{r}) = \mathcal{F}^{-1}(T(\boldsymbol{g}))$$

$T(\boldsymbol{g})$ is given by:

$$T(\boldsymbol{g}) = A(\boldsymbol{g})exp(i\chi(\boldsymbol{g}))$$

where $A(\boldsymbol{g})$ is the aperture function and the phase shift $\chi(\boldsymbol{g})$ is related to a set of microscope parameters including the acceleration voltage, the semi-convergence angle, defocus and spherical aberration of the third and fifth order. The microscope parameters used to generate $\chi(\boldsymbol{g})$ are listed in Table S3.

**Table S3 Parameters to generate phase shift $\chi(g)$**

| Parameter | Value |
|---|---|
| Acceleration voltage | 300 kV |
| Defocus | Half of column thickness |
| Spherical aberration | 0 mm |
| Spherical aberration of 5th order | 0 mm |
| Semi-convergence angle | 20 mrad |

The probe intensity $psf(\boldsymbol{r})$ is then convolved with the projected potential $P(\boldsymbol{r})$:

$$O(\boldsymbol{r}) = P(\boldsymbol{r}) * psf(\boldsymbol{r})$$

followed by fitting $O(\boldsymbol{r})$ with 2D Lorentzian or 2D Gaussian peaks written as:

$$f_{kl}^{Lorentzian}(\theta) = \eta(\rho^2 + (x_k - \beta_x)^2 + (y_k - \beta_y)^2)^{-\frac{3}{2}}$$

$$f_{kl}^{Gaussian}(\theta) = \eta \exp\left[\frac{-(x_k - \beta_x)^2 - (y_l - \beta_y)^2}{2\rho^2}\right]$$

with $f_{kl}$ the pixel intensity at pixel (k, l) in the model and $\theta = (\beta_x, \beta_y, \eta, \rho)^T$ the vector of unknown parameters, where $\rho$, $\beta_x$ and $\beta_y$ are the width, and x- and y- coordinates of the peak. The height of the peak is $\frac{\eta}{\rho^3}$ for the 2D Lorentzian and $\eta$ for the 2D Gaussian. Use has been made of the uniformly weighted least squares criterion, which evaluates the correspondence between $O(r)$ and the model. The estimates $\hat{\theta}$ are given by the values of $t$ that minimize the uniformly weighted least square criterion:

$$\hat{\theta} = \arg\min_{t} \sum_{k=1}^{K} \sum_{l=1}^{L} \left[ \left( O_{kl} - f_{kl}^{Lorentzian(Gaussian)}(t) \right)^2 \right]$$

where $O_{kl}$ represent the observed values in $O(r)$ at pixel $(k, l)$.